\documentclass[preprint,12pt]{elsarticle}
\usepackage{amssymb}
\usepackage{amsmath}
\usepackage{algorithm}
\usepackage{algorithmic}
\usepackage{ulem}
\usepackage{multirow}
\usepackage{amsfonts,amssymb,amsbsy,textcomp,marvosym,picins,amsmath,caption,threeparttable,amsthm,subfigure}
\usepackage{eurosym,mathrsfs,fancyhdr,CJK,multicol,graphics,indentfirst,color,bm,upgreek,booktabs,graphicx}
\usepackage{enumerate}
\journal{Elsevier}

\begin{document}

\begin{frontmatter}

\title{Alleviating the recommendation bias via rank aggregation}

\author[UESTC]{Qiang Dong\corref{cor1}}
\cortext[cor1]{Corresponding author}
\ead{dongq@uestc.edu.cn}
\author[UESTC]{Quan Yuan}
\author[UESTC]{Yang-Bo Shi}

\address[UESTC]{School of Computer Science and Engineering, University of Electronic Science and Technology of China, Chengdu 611731, China}

\begin{abstract}
The primary goal of a recommender system is often known as "helping
users find relevant items", and a lot of recommendation algorithms are proposed accordingly. However, these accuracy-oriented methods usually suffer the problem of recommendation bias on popular items, which is not welcome to not only users but also item providers. To alleviate the recommendation bias problem, we propose a generic rank aggregation framework for the recommendation results of an existing algorithm, in which the user- and item-oriented ranking results are linearly aggregated together, with a parameter controlling the weight of the latter ranking process. Experiment results of a typical algorithm on two real-world data sets show that, this framework is effective to improve the recommendation fairness of any existing accuracy-oriented algorithms, while avoiding significant accuracy loss.

\end{abstract}


\end{frontmatter}

\section{Introduction}
\label{sec-Introduction}

Nowadays, as an effective way to relieve the information
overload problem confronting us, recommender systems have achieved
great development in the recent decades \cite{kbs-survey,ZhaoYD,ZhangZK1,RS,ZhangZK2}. When people feel
overwhelmed with a huge collection of online items, such as
e-books, movies and music, recommender systems arise to help them
find the relevant ones from seemingly countless candidates,
by learning from the past behaviors of these people. To some extent,
recommender systems take the place of the users themselves to assess
the relevance of items, including tweets, news, restaurants, and
even online social friends.

Most of early recommendation systems put great emphasis on the
accuracy of recommendation results. For example, the Netflix
corporation used to offer a reward of one million US Dollars to
improve the recommendation accuracy of video watching on the website
netflix.com \cite{Xiang}. These systems more or less suffer the problem of recommendation bias to popular items, that is, recommending a few popular items to a majority of users. This will prevent the users from exploring more personalized items, and thus decrease user's loyalty and trust in the recommender system. Until the recent decade, the diversity of recommendation results, which is introduced to solve the above-mentioned problem, begins to receive significant research attentions \cite{kbs-survey,RS,ZhuCai}.

However, both of accuracy and diversity are introduced to evaluate
recommender systems from the standpoint of users. Recently, Jannach
and Adomavicius \cite{purpose} proposed to revisit the purposes and goals of recommender systems from the item-provider viewpoints. One major concern of the item providers is how to increase the sales of long-tailed items, i.e., the providers wish that the algorithms could produce recommendation results with uniform coverage on all items in the system, especially the long-tailed ones \cite{Learning-to-Rank,Lv-congestion}. Vargas and Castells \cite{inverted} proposed the reversed-neighborhood Collaborative Filtering algorithms, which exchange the roles of users and items, and then compute the recommendation scores of all the user-item pairs following the original procedure. The experiment results show that, the reversed recommendation methods offer better coverage and diversity, but the accuracy may suffer some loss.

Taking into account the coverage/diversity advantage and accuracy disadvantage of reversed recommendation, we consider to aggregate the direct and reversed rank results of some algorithm together, with the hope of greatly improving the coverage and diversity, while keeping the accuracy unchanged or slightly increased. The main contributions of this paper are three-folded.

\begin{enumerate}[(1)]

\item We empirically find that in the recommendation results of P3 algorithm, the direct rank number of a user-item pair shows a negative correlation with the item degree, while the reversed rank numbers with the same item degree are almost uniformly distributed in the whole range.

\item We propose the Two-Way Rank Aggregation framework of the direct and reversed rank results, which is a re-sorting method and can be combined with any existing recommendation algorithm.

\item We apply the proposed aggregation framework in the famous P3 algorithm, and find that, our proposed framework can produce recommendation results with superior accuracy, coverage and diversity, which are also comparable to state-of-the-art P3 improved algorithms.

\end{enumerate}

\section{Preliminary knowledge}

The topological structure of a recommender system is usually represented as a bipartite network, consisting of two types of nodes, $m$ user nodes and $n$ item nodes, and the links between them. For convenience, we use $u$, $v$ or $u_k$ to denote a \emph{u}ser, $1 \leq k \leq m$, $i$, $j$ or $i_t$ an \emph{i}tem, $1 \leq t \leq n$, and $(u,v)$ or $(u_k,i_t)$ a user-item link. For a user node $u$, its neighborhood is defined by all the item nodes connected to $u$, denoted as $I(u)$, and its degree is the cardinality of $I(u)$, denoted as $k_u$ or $k(u)$. The counterparts for an item node $i$ are $U(i)$ and $k_i$ or $k(i)$.

This bipartite network could be fully described by an adjacency matrix $A = \{a_{kt}\}_{m\times n}$, where the element $a_{kt} = 1$ if there exists a link between user $u_k$ and item $i_t$ (user $u_k$ collects item $i_t$), meaning that user $u_k$ declared explicitly his preference on item $i_t$ in the past, and $a_{kt} = 0$ otherwise. For every target user, the essential task of a recommender system becomes to recommend him a ranked sublist of uncollected items of his potential interest.

\subsection{Data sets}

Two benchmark data sets are employed to evaluate the performance of recommendation algorithms, namely, Movielens and Netflix. Both of them are movie rating data set, where users rate their watched movies (rephrased as items in this paper) with an explicit integer scores from 1 to 5. For each data set, we use only the ratings no less than 3 to construct links of the bipartite network. Table 1 summarizes the statistical features of the two data sets. To evaluate the offline performance of different models, each data set is randomly divided into two subsets: the training set containing $90\%$ of the links and the probe set $10\%$ of the links. The training set is treated as known information to
make recommendation and the probe set is used to test the accuracy performance of the recommendation results.

\begin{table*}
\caption{Basic statistics of real-world networks used in this paper, including the number of users, item and links, and the sparsity that is the proportion of links to the total number of possible links.}
\label{tab:01}
\begin{center}
\begin{tabular}[b]{lllll }
\hline
Data set & \#User & \#Items & \#Links & Sparsity \\
\hline
Movielens & 6,000 & 3,600 & 800,000 & 3.8\%  \\
Netflix & 9,500 & 14,000 & 1,700,000 & 1.2\% \\
\hline
\end{tabular}
\end{center}
\end{table*}

\subsection{Evaluation measures}

In order to evaluate the performance of recommendation algorithms, we adopt three typical metrics in this paper, which can be categorized into accuracy, diversity and coverage measures.

When designing a recommender algorithm, one major concern is the accuracy of recommendation results. We make use of Precision to measure the recommendation accuracy. For a target user $u$, the recommender system will return him a ranked list of his uncollected items. Precision is the fraction of accurately recommended items to the length of recommendation lists. By averaging over the Precision of all users, we obtain the overall Precision of the system,

\begin{center}
$\text{Precision}(L) = \frac{1}{m} \sum\limits_{k=1}^m \frac{h_k}{L}$,
\end{center}
where $L$ is the fixed length of recommendation list, $m$ the total number of users in the system, and $h_k$ the number of accurately recommended items in the recommendation list of $u_k$.



Besides accuracy, diversity has been regarded as another important concern in evaluating a recommendation model. The diversity measures how different are the recommendation lists  from each other, and can be quantified by Hamming Distance. Given two different users $u$ and $v$, borrowing inspiration from the Hamming distance of two strings, we can calculate recommendation diversity in a similar way, as

\begin{center}
$\text{HD}_{uv}(L)=1-\frac{D_{uv}(L)}{L}$
\end{center}
where $D_{uv}(L)$ is the number of common items in the recommendation lists of length $L$ of users $u$ and $v$. Clearly, the higher is the value, the more personalized lists are recommended to users. By averaging over the Hamming Distances of all user pairs, we obtain the overall diversity of the system.

Coverage is usually measured by the percentage of items
that an algorithm is able to recommend to users. While being simple and intuitive, this definition may not be very robust, since the contribution to the metric of an item that has been recommended just once is equal to that of other item recommended a thousand times \cite{inverted}. Therefore, Fleder and Hosanagar \cite{Gini} proposed to use the Gini coefficient to measure the imbalance in the number of times each item is recommended,

\begin{center}
$\text{Gini}(L)=1-\frac{1}{n-1}\sum\limits_{q=1}^n {(2q-n-1)p(i_q)}$
\end{center}
where $i_q$ is the $q$-th least recommended item, $p(i_q)$ is the ratio of the recommended times of $i_q$ to the recommended times of all the items.

\subsection{Baseline algorithms}

For every user-item pair, say user $u$ and item $i$, an algorithm $\pi$ will produce a score, $\pi\text{-score}(u,i)$, predicting how much user $u$ likes item $i$.
Next, we briefly review the famous $P^3$ algorithm and three of its improved versions.

\begin{enumerate}[(1)]
\item The $P3$ algorithm, also known as NBI \cite{NBI} or ProbS \cite{H&P} algorithm, can be seen as a three-step random walk process in the user-item bipartite network. The walker starts on the target user $u$, and at each step moves to randomly chosen one of its neighbor nodes, finally arrives the target item $i$ after three steps. The transition probability of the walker from current node to its neighbor node is equal to the reciprocal of the degree of current node. Therefore, the transition probability from $u$ to $i$ is
\begin{center}
$P3\text{-score}(u,i)=\sum\limits_{j=1}^{n}\sum\limits_{v=1}^{m}{\frac{a_{uj}a_{vj}a_{vi}}{k_{u}k_{j}k_{v}}}$.
\end{center}
Since the transition probability of the first step is identical for the same user, the divisor $k_{u}$ is deleted from the above formula in the remainder of this paper.



\item The $P^3_\alpha$ algorithm \cite{P3}, raises the transition probability of $P^3_\alpha$ to the power of $\alpha$, i.e.,

\begin{center}
$P_\alpha^3\text{-score}(u,i)=\sum\limits_{j=1}^{n}\sum\limits_{v=1}^{m}{\frac{a_{uj}a_{vj}a_{vi}}{(k_{j}k_{v})^\alpha}}$
\end{center}
where the parameter $\alpha$ ranges over the real numbers. we can see that when $0 < \alpha < 1$, the transition probability of every step is zoomed in, and $\alpha > 1$ zoomed out. However, the zoom scale of unpopular items is much smaller than that of popular ones.

\item The $RP_\beta^3$ algorithm \cite{RP3}, introduces a simple re-ranking procedure dependent on the popularity of target item into the score of $P^3$ as follows,

\begin{center}
$RP_\beta^3\text{-score}(u,i)=\frac{1}{k_i^\beta}
\sum\limits_{j=1}^{n}\sum\limits_{v=1}^{m}{\frac{a_{uj}a_{vj}a_{vi}}{k_{j}k_{v}}}$
\end{center}
where the parameter $\beta$ ranges over the non-negative numbers. we can see that if $\beta = 0$, it reduces to $P3$. When $\beta > 0$, the popular items will get smaller scores, and the larger is $\beta$, the stronger is the score suppression of popular items.

\item  The $HHP$ algorithm \cite{H&P}, is a nonlinear hybrid of ProbS and HeatS, which is written as,
\begin{center}
     $HHP\text{-score}(u,i)=
\sum\limits_{j=1}^{n}\frac{a_{uj}}{k_i^{1-\lambda}k_{j}^\lambda}\sum\limits_{v=1}^{m}{\frac{a_{vj}a_{vi}}{k_{v}}}$
\end{center}
where the parameter $\lambda$ ranges from 0 to 1. when $\lambda = 1$, it reduces to ProbS or $P3$, and when $\beta = 0$, it reduces to HeatS.

\end{enumerate}

Compared with the original $P3$ algorithm, all the three improved ones can greatly increase the recommendation diversity and accuracy, of course to different extents \cite{H&P,P3,RP3}. \textcolor[rgb]{1.00,0.00,0.00}{Besides the above listed three algorithms, P3 is also improved in many other ways \cite{NieDC,LiWJ,AnYH,Zhu-diffusion-1,Zhu-diffusion-2}. The interested reader is referred to a survey article for comprehensive comparisons \cite{ZengAn}.}

\section{Main results}

As we know, many well-known recommendation algorithms more or less suffer the problem of popularity bias on recommended items. Taking $P3$ algorithm as an example (see Figure 1), we can see that the degree distribution of items in two data sets is long-tailed, while the degree distribution of recommended items produced by $P3$ algorithm differs greatly from that in original systems. The unpopular ones comprising the majority of items are rarely recommended, while a small number of popular items are predominant in the recommendation lists. From the point of view of providers, this is the last thing they want to see. Undoubtedly, providers prefer recommendation results with more uniform coverage on all items in the system, especially the upopular ones.

\setcounter{figure}{0}
\begin{figure*}[!htb]
\centering
  \subfigure[]{
    \includegraphics[width=2.5in]{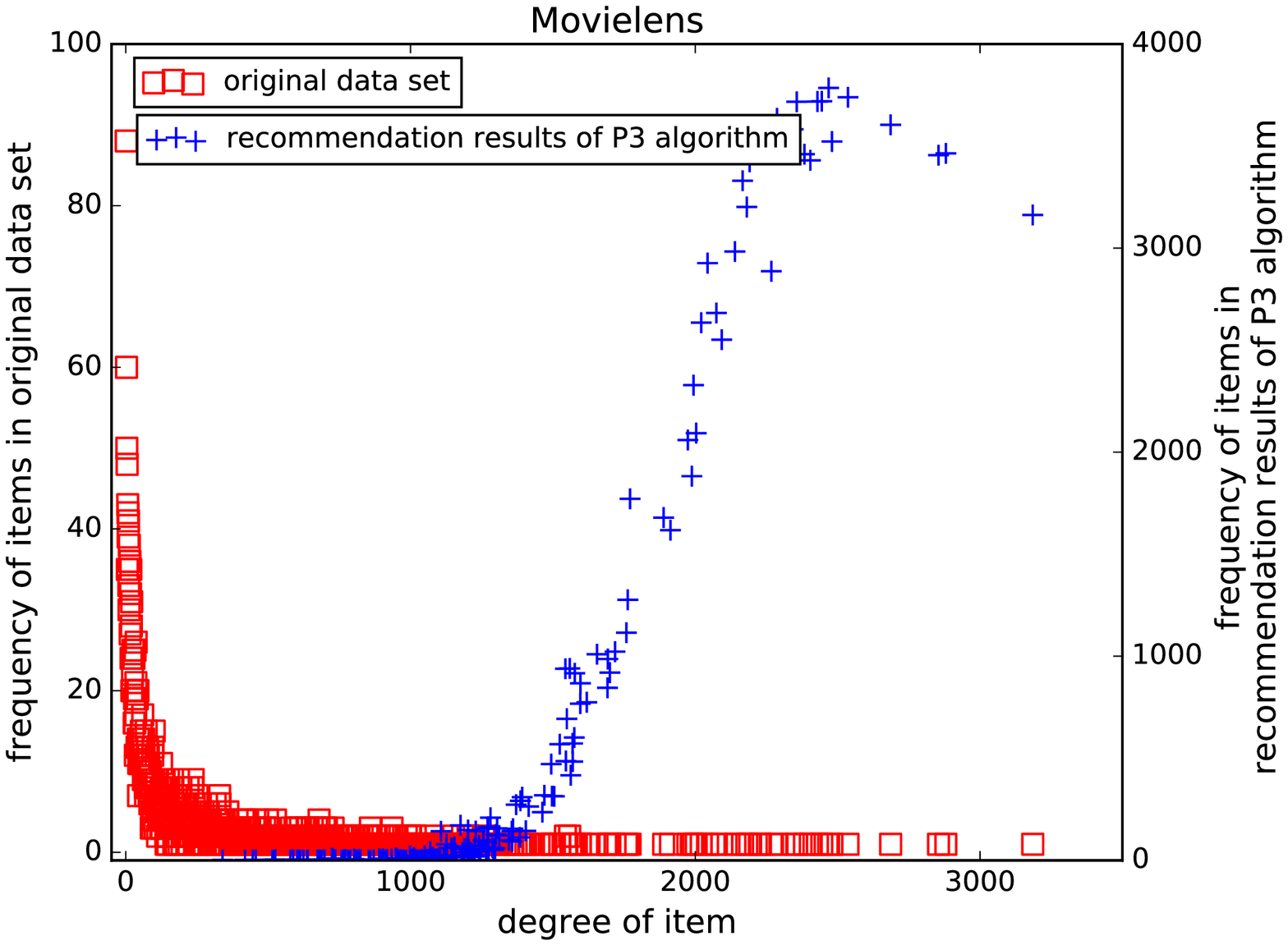}}
  \subfigure[]{
    \includegraphics[width=2.5in]{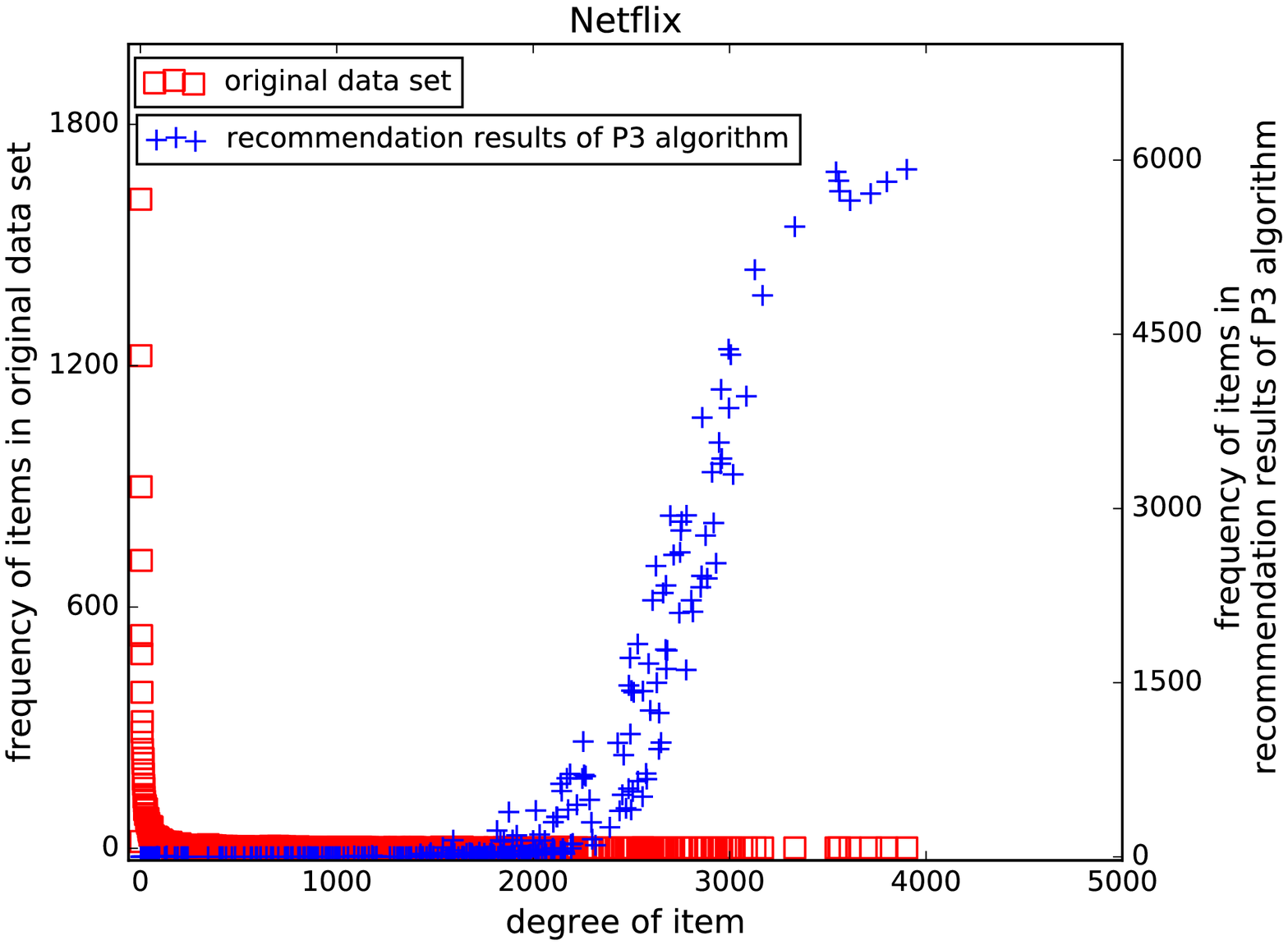}}
  \caption{The degree distribution of items in original (a) Movielens and (b) Netflix data sets and recommendation lists of $P3$ algorithm on the same data set. The left ordinate is the frequency of items of specific degree in original data set, and the right ordinate is the recommendation frequency of items of specific degree in recommendation results of $P3$ algorithm.}
\end{figure*}

The reversed recommendation \cite{inverted} is a way to address this coverage problem, but the accuracy may suffer some loss.
Since the direct recommendation usually enjoys high recommendation accuracy yet low coverage, why don't we aggregate the direct and reversed rank results together to get higher recommendation coverage and accuracy?

%
%

\subsection{The rank aggregation framework}

Given a user $u$, we rank all his uncollected items in descending order of $\pi\text{-score}(u,i_t)$, $t\in\{1,2,\cdots, n\}$, and refer to the rank number of item $i_t$ in this sequence as the forward rank number, denoted by $\text{f-rank}^\pi(u,i_t)$. Given an item $i$, all the users who do not collect it are ranked in descending order of $\pi\text{-score}(u_k,i)$, $k\in\{1,2,\cdots, m\}$, then we can similarly define the backward rank number $\text{b-rank}_\pi(u_k,i)$.
In short, a specified user-item pair is corresponding to two different rank numbers, f-rank number and b-rank number, respectively from the viewpoints of user and item.

\setcounter{figure}{1}
\begin{figure*}[!htb]
\centering
  \subfigure[]{
    \includegraphics[width=2.5in]{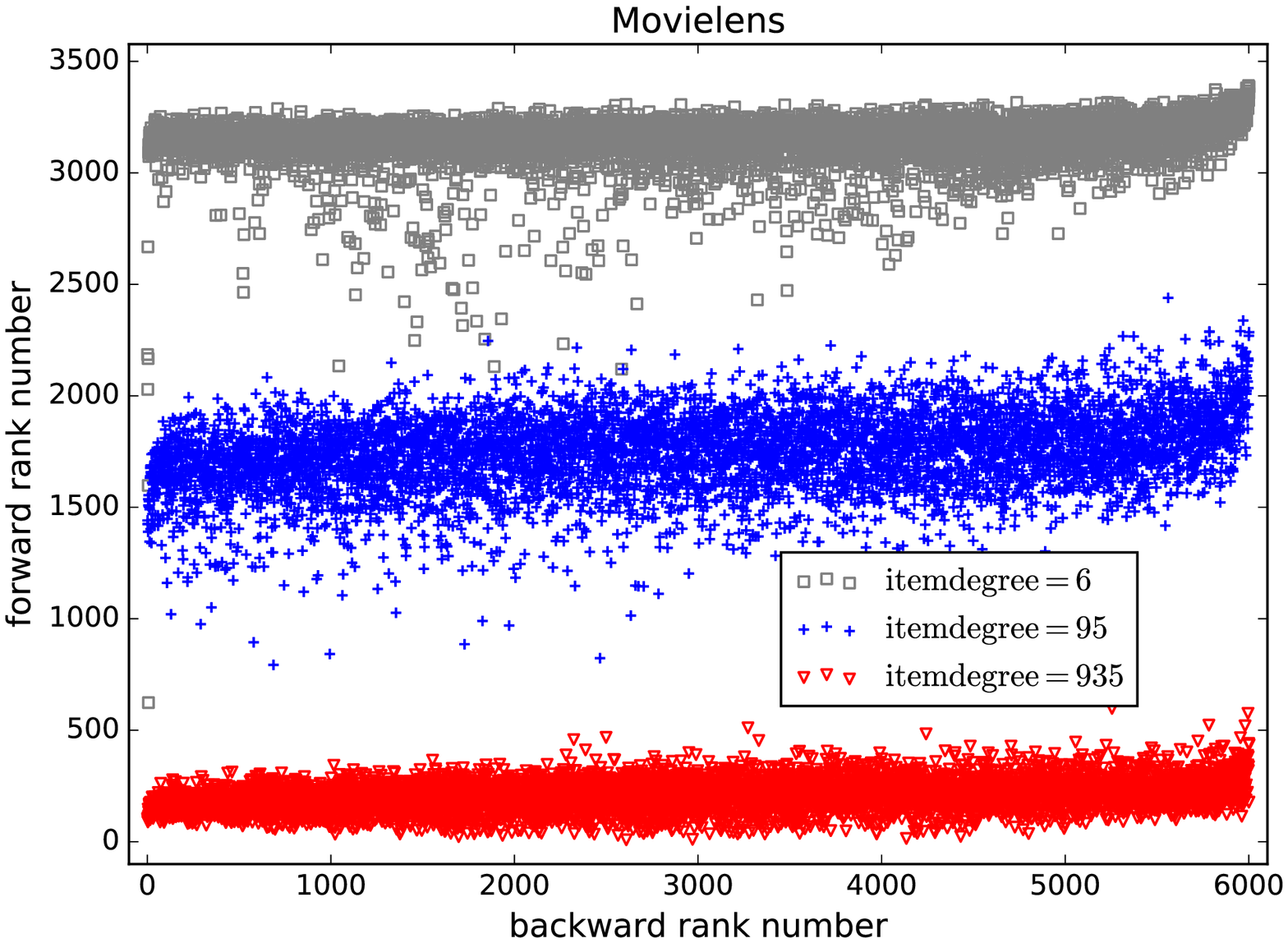}}
  \subfigure[]{
    \includegraphics[width=2.5in]{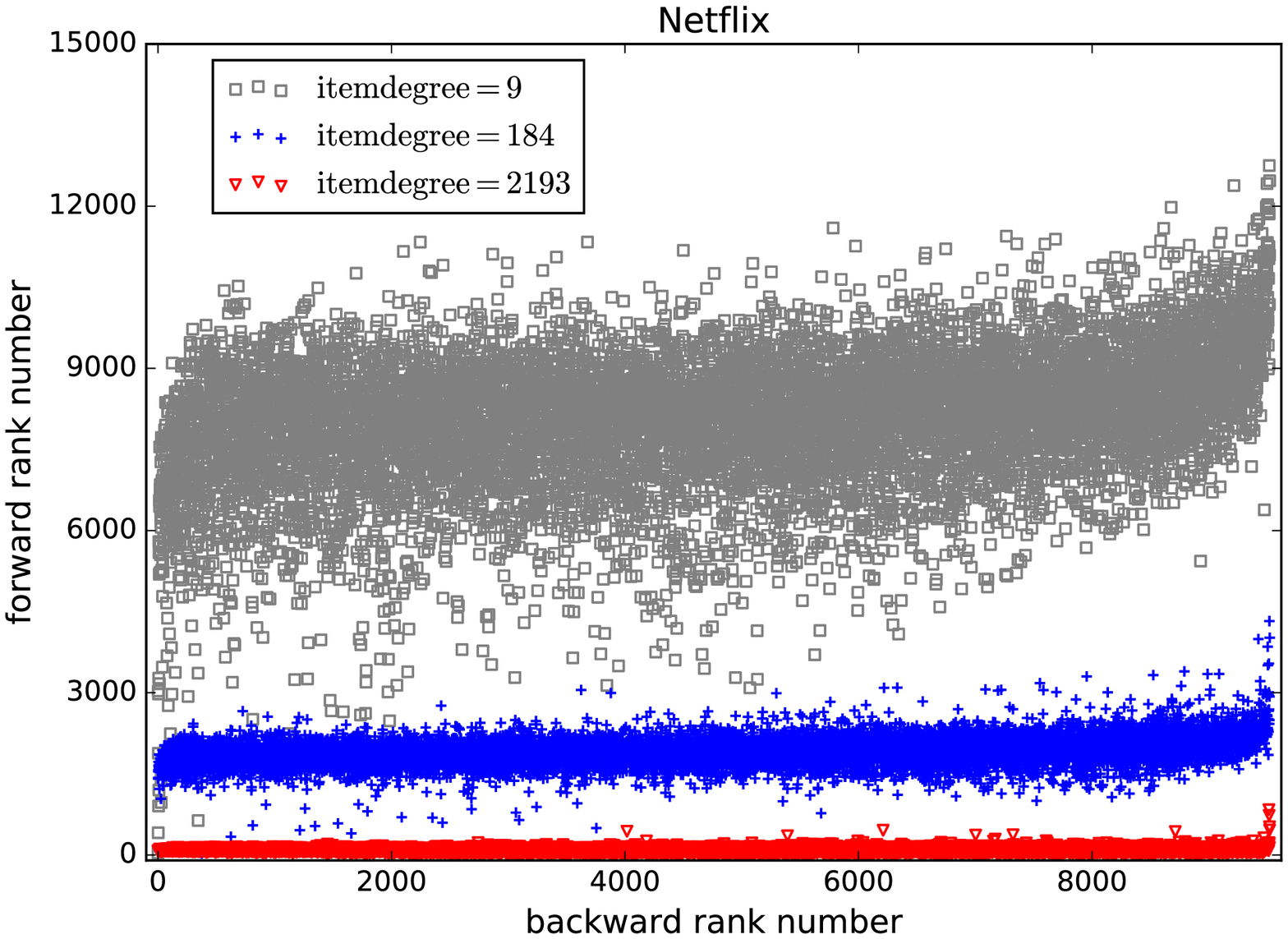}}
  \caption{The scatter plot of the forward rank number against the backword rank number of the same user-item pair in (a) Movielens and (b) Netflix data sets, respectively for three manually chosen values of small, median and large item degree.}
\end{figure*}

Following the results of $P3$ algorithm on Movielens and Netflix data sets, we present two scatter plots of the f-rank number against the b-rank number of the same user-item pair in Figure 2, respectively for three manually chosen values of small, median and large item degree. It is observed that the f-rank number exhibits a negative correlation with the item degree. This means that most unpopular items seldom get the opportunity to be recommended to users, and a few popular items always appear in the head of recommendation lists. That is why the recommendation coverage of P3 algorithm is not that good. Different from the f-rank numbers, the b-rank numbers for three values of item degree, are almost uniformly distributed in the whole range, which is preferred by a personalized recommender system.

Based on this observation, we introduce the aggregation rank number, which is a weighted linear aggregation of the forward and backward rank numbers, defined by
\begin{center}
$\text{ag-rank}^\pi(u,i)=(1-\lambda)*\text{f-rank}^\pi(u,i)+\lambda *
\text{b-rank}^\pi(u,i)$,
\end{center}
where $\lambda$ is an adjustable parameter ranging in closed interval [0,1].
Finally, we give a target user a recommendation list of his uncollected items in ascending order of this aggregation rank number. We call this re-ranking framework as Two Way Ranking Aggregation, shorted by TWRA, and accordingly denote the algorithm $\pi$ combined with this framework as TWRA($\pi$). when $\lambda = 0$, the framework reduces to the original algorithm $\pi$. while $\lambda = 1$, the framework will produce recommendation results merely from the standpoint of items, regardless of  user experience.

\subsection{Performance analysis}

\setcounter{figure}{2}
\begin{figure*}[!htb]
\centering
  \subfigure[]{
    \includegraphics[width=2.5in]{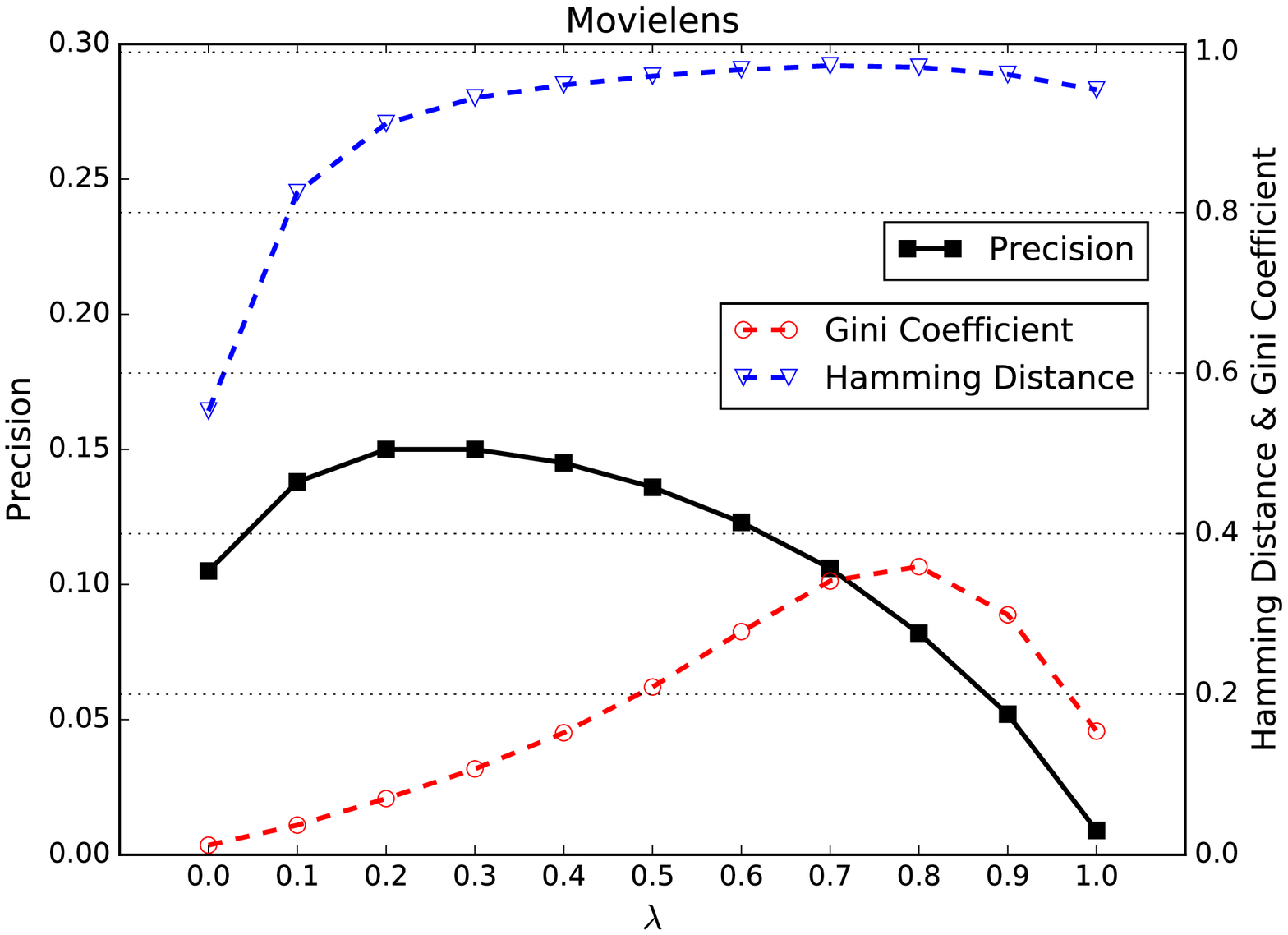}}
  \subfigure[]{
    \includegraphics[width=2.5in]{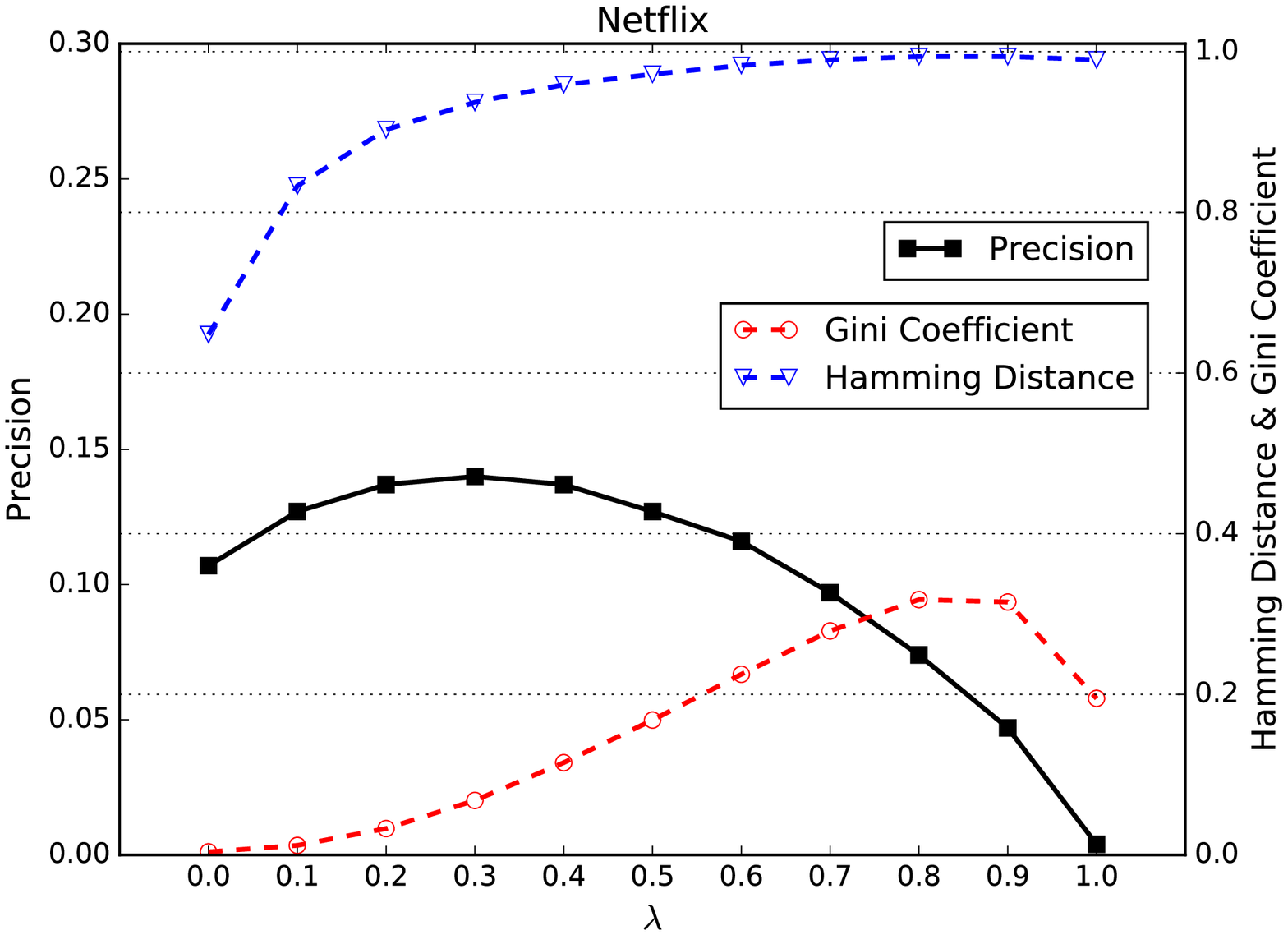}}
  \caption{The performance of TWRA($P^3$) model against the parameter $\lambda$ on (a) MovieLens and (b) Netflix data sets. The left ordinate is for the Precision value, and the right ordinate is for the Hamming Distance and Gini Coefficient.}
\end{figure*}

Figure 3 plots the performance of TWRA($P^3$) model against the parameter $\lambda$ on MovieLens and Netflix data sets, respectively. The length of recommendation list is set to 20. We find that as the parameter $\lambda$ is traversed from 0 to 1, the values of diversity and coverage increase in most of the range, and fall down when $\lambda$ approaches 1. The curves of accuracy have a slightly rising head and a long dropping tail, where the $\lambda$ value for the break point is closer to 0 than 1 on both data sets. In a word, the coverage/diversity values and the accuracy values become a trade-off in most range of $\lambda$, with a short coincident increase when $\lambda$ is small. Therefore, we empirically choose the optimal $\lambda$ value to be 0.3 and 0.4 respectively on Movielens and Netflix data sets in the following performance comparison.

\begin{table*}
\small
\caption{Performance of all algorithms on MovieLens and Netflix data sets. Parameterized algorithms are represented by parameter value resulting in second-best precision performance. The best value of each metric is highlighted in bold. The length of recommendation list is 20.}
\label{tab:comparison}
\begin{center}
\begin{tabular}[b]{p{1.8cm}p{4cm}ccc }
\hline
Data set &   Method&  Precision  &  Hamming Distance &   Gini Coefficient\\
\hline
\multirow{5}*{MovieLens}& $P3$ &          0.105      &  0.553   &                   0.012 \\
~& $P^3_\alpha$, $\alpha = 1.7$ &  0.105      &  0.837   &                   0.082 \\
~& $RP^3_\beta$, $\beta = 0.7$ &  \textbf{0.155}      &  0.885   &           0.091\\
~& $HHP_\lambda$, $\lambda = 0.3$ & 0.149      &  0.853 &                   0.057 \\
~& TWRA($P3$), $\lambda = 0.3$ &   0.150      &  \textbf{0.943} &  \textbf{0.107} \\
\hline
\multirow{5}*{Netflix}& $P3$ &          0.107 &      0.648   &                      0.004 \\
~&$P^3_\alpha$, $\alpha = 1.4$ &  0.101 &      0.842   &                      0.027 \\
~&$RP^3_\beta$, $\beta = 0.6$ &  0.127 &      0.877   &                      0.068 \\
~&$HHP_\lambda$, $\lambda = 0.2$ & 0.129 &      0.822 &                      0.015 \\
~&TWRA($P3$), $\lambda = 0.4$ &   \textbf{0.137} &   \textbf{0.959} &   \textbf{0.115} \\

\hline
\end{tabular}
\end{center}
\end{table*}

Table 2 summarizes the detailed values of evaluation measures on two data sets when the length of recommendation list is set to 20. Compared with the original $P3$ algorithm, TWRA($P3$) model greatly improves the values of Hamming Distance and Gini coefficient by 70.5\% and 792\% on Movielens data set, and 48\% and 2875\% on Netflix data set. What is more, TWRA($P3$) model also improves the accuracy value by 42.6\%. The significant improvements on Gini coefficient again confirm the disadvantage of popularity bias of original $P3$ algorithm.

According to Table 2, TWRA($P3$) model beats state-of-the art P3 improved algorithms on Hamming Distance and Gini coefficient, while its accuracy is also comparable.
Compared with the second-best $RP^3_\beta$ algorithm, the increased percentage of TWRA($P3$) on diversity and coverage is much larger on Netflix than Movielens data set, and
the accuracy is better on Netflix data set but a little worse on Movielens data set. Thus, we get the conclusion that
TWRA framework play a more significant role in sparser (Netflix) than denser (Movielens) data set. The possible reason is that
the sparser is the data set, the information that can be utilized by recommendation algorithm is less, where the reversed rank will provide useful supplement to the direct recommendation.
Since most of current commercial recommender systems are deployed on sparse user-item networks, our proposed rank aggregation framework is of great practical significance.

\section{Concluding remarks}

\textcolor[rgb]{1.00,0.00,0.00}{A basic but general solution of trading off accuracy and coverage/diversity is weighted linear aggregation of an accuracy-oriented recommender and a diversity-oriented recommender. Though easy to implement, this approach has the disadvantage of requiring two independent recommendation calculations, thus increasing computational cost \cite{P3}. We proposed a rank aggregation framework, by running the algorithm only once, getting the regular predictive scores, and combining the item- and user-oriented ranking numbers together into the final ranking number. This framework is effective to improve the coverage/diversity of any existing accuracy-oriented algorithms, while avoiding significant accuracy loss.}


\section*{Acknowledgements}

The authors would like to express their gratitude to the editor and the three anonymous referees for their valuable suggestions that have greatly improved the quality of the paper.

This work is supported by UESTC Fundamental Research Funds for the Central Universities (Grant No. ZYGX2016J196).

\end{document}